\documentclass[a4paper,11pt,showkeys]{article}
\usepackage{amsmath,amssymb,amsfonts}
\usepackage[numbers,sort&compress,merge]{natbib}
\usepackage{graphicx}
\usepackage{rotating}
\usepackage[hang,bf,nooneline]{subfigure}
\usepackage[hang,bf,nooneline]{caption}

\newlength{\dslashwidth}

\newcommand{\tb}{\ensuremath{\tan\beta}}

\newcommand{\beq}{\begin{equation}} 
\newcommand{\eeq}{\end{equation}}
\newcommand{\beqa}{\begin{eqnarray}} 
\newcommand{\eeqa}{\end{eqnarray}}
\newcommand{\newc}{\newcommand}
\newcommand{\bq}{\begin{equation}}
\newcommand{\eq}{\end{equation}}
\newcommand{\ba}{\begin{array}}
\newcommand{\ea}{\end{array}}
\newcommand{\bqa}{\begin{eqnarray}}
\newcommand{\eqa}{\end{eqnarray}}

\newcommand{\lnf}{{\ifmmode \Lambda^{(N_f)} \else $\Lambda^{(N_f)}$\fi}}
\newcommand{\ms}{{\ifmmode \overline{MS} \else $\overline{MS}$\fi}}
\newcommand{\dr}{{\ifmmode \overline{DR} \else $\overline{DR}$\fi}}
\newcommand{\lms}{{\ifmmode \Lambda^{(5)}_{\overline{MS}} \else $\Lambda^{(5)}_{\overline{MS}}$\fi}}
\newcommand{\lam}{{\ifmmode \Lambda \else $\Lambda$\fi}}
\newcommand{\mev}{{\ifmmode {\rm MeV} \else ${\rm MeV}$\fi}}
\newcommand{\gev}{{\ifmmode {\rm GeV} \else ${\rm GeV}$\fi}}
\newcommand{\gevc}{{\ifmmode {\rm GeV/c^2} \else ${\rm GeV/c^2}$\fi}}
\newcommand{\tev}{{\ifmmode {\rm TeV} \else ${\rm TeV}$\fi}}
\newcommand{\tevc}{{\ifmmode {\rm TeV/c^2} \else ${\rm TeV/c^2}$\fi}}
\newcommand{\lp}{{\ifmmode L^+  \else $L^+$\fi}}
\newcommand{\lm}{{\ifmmode L^-  \else $L^-$\fi}}
\newcommand{\mlp}{{\ifmmode M(L^-) \else $M(L^-)$\fi}}
\newcommand{\mlz}{{\ifmmode M(L^0) \else $M(L^0)$\fi}}
\newcommand{\lz}{{\ifmmode L^0 \else $L^0$\fi}}
\newcommand{\ev}{{\ifmmode GeV/c^2 \else $GeV/c^2$\fi}}
\newcommand{\tri}{{\ifmmode \triangleup \else $\triangleup$\fi}}
\newcommand{\unl}{{\ifmmode U_{lL^0} \else $U_{lL^0}$\fi}}\newcommand{\gL}{{\ifmmode g_L \else $g_{L}$\fi}}
\newcommand{\gR}{{\ifmmode g_R  \else $g_{R}$\fi}}
\newcommand{\gumu}{{\ifmmode \gamma^{\mu} \else $\gamma^{\mu}$\fi}}
\newcommand{\gunu}{{\ifmmode \gamma^{\nu} \else $\gamma^{\nu}$\fi}}
\newcommand{\gdmu}{{\ifmmode \gamma_{\mu} \else $\gamma_{\mu}$\fi}}
\newcommand{\gdnu}{{\ifmmode \gamma_{\nu} \else $\gamma_{\nu}$\fi}}
\newcommand{\stw}{{\ifmmode\sin^2\theta_W \else $\sin^{2}\theta_{W}$ \fi}}
\newcommand{\sws}{{\ifmmode \;\sin^2\theta_W  \else $\;\sin^{2}\theta_{W}$ \fi}}
\newcommand{\cws}{{\ifmmode \;\cos^2\theta_W  \else $\;\cos^{2}\theta_{W}$ \fi}}
\newcommand{\sw}{{\ifmmode \;\sin\theta_W  \else $\sin\theta_{W}$ \fi}}
\newcommand{\cw}{{\ifmmode \;\cos\theta_W  \else $\;\cos\theta_{W}$ \fi}}
\newcommand{\tw}{{\ifmmode \;\tan\theta_W  \else $\;\tan\theta_{W}$ \fi}}
\newcommand{\qq}{{\ifmmode q\overline{q} \else $q\overline{q}$\fi}}
\newcommand{\lR}{{\ifmmode l_R \else $l_R$\fi}}
\newcommand{\lL}{{\ifmmode l_L \else $l_L$\fi}}
\newcommand{\nt}{{\ifmmode \nu_{\tau} \else $\nu_{\tau}$\fi}}
\newcommand{\nuR}{{\ifmmode \nu_R  \else $\nu_R$\fi}}
\newcommand{\nuL}{{\ifmmode \nu_L  \else $\nu_L$\fi}}
\newcommand{\qR}{{\ifmmode g_R  \else $q_R$\fi}}
\newcommand{\qL}{{\ifmmode q_L  \else $q_L$\fi}}
\newcommand{\qRp}{{\ifmmode q_R'  \else $q_{R}$'\fi}}
\newcommand{\qLp}{{\ifmmode q_L'  \else $q_{L}$'\fi}}
\newcommand{\est}{{\ifmmode e^{\bf \ast} \else $e^{\bf \ast}$\fi}}
\newcommand{\lst}{{\ifmmode l^{\bf \ast} \else $l^{\bf \ast}$\fi}}
\newcommand{\must}{{\ifmmode \mu^{\bf \ast} \else $\mu^{\bf \ast}$\fi}}
\newcommand{\taust}{{\ifmmode \tau^{\bf \ast} \else $\tau^{\bf \ast}$ \fi}}
\newcommand{\pperp}{{\ifmmode p_t  \else $p_t$\fi}}
\newcommand{\et}{{\ifmmode E_t  \else $E_t$\fi}}
\newcommand{\xt}{{\ifmmode x_t  \else $x_t$\fi}}
\newcommand{\smumu}{{\ifmmode \sigma_{\mu\mu}  \else $\sigma_{\mu\mu}$ \fi}}
\newcommand{\eg}{{\ifmmode e\gamma  \else $e\gamma$\fi}}
\newcommand{\epem}{{\ifmmode e^+e^-  \else $e^+e^-$\fi}}
\newcommand{\lplm}{{\ifmmode L^+L^-  \else $L^+L^-$\fi}}
\newcommand{\pp}{{\ifmmode p\overline p  \else $p\overline p$\fi}}
\newcommand{\llz}{{\ifmmode L^0\overline{L}^0 \else $L^0\overline{L}^0$\fi}}
\newcommand{\epemt}{{\ifmmode e^+e^- \to  \else $e^+e^- \to$\fi}}
\newcommand{\eb}{{\ifmmode E_{beam}  \else $E_{beam}$\fi}}
\newcommand{\ip}{{\ifmmode pb^{-1}  \else $pb^{-1}$\fi}}
\newcommand{\upm}{{\ifmmode ^{\pm}  \else $^{\pm}$\fi}}
\newcommand{\de}{{\ifmmode ^{\circ}  \else $^{\circ}$ \fi}}
\newcommand{\appr}{{\ifmmode \sim \else $\sim$ \fi}}
\newcommand{\corresp}{{\ifmmode \stackrel{\wedge}{=} \else $\stackrel{\wedge}{=}$ \fi}}
\newcommand{\sqrts}{{\ifmmode \sqrt{s} \else $\sqrt{s}$\fi}}
\newcommand{\zz}{{\ifmmode Z^0  \else $Z^0$\fi}}
\newcommand{\mz}{{\ifmmode M_{Z}  \else $M_{Z}$\fi}}
\newcommand{\mzs}{{\ifmmode M_{Z}^2  \else $M_{Z}^2$\fi}}
\newcommand{\mw}{{\ifmmode M_{W}  \else $M_{W}$\fi}}
\newcommand{\mws}{{\ifmmode M_{W}^2  \else $M_{W}^2$\fi}}
\newcommand{\mh}{{\ifmmode M_{Higgs}  \else $M_{Higgs}$\fi}}
\newcommand{\gt}{{\ifmmode \Gamma_{tot} \else $\Gamma_{tot}$\fi}}
\newcommand{\msusy}{{\ifmmode M_{SUSY}  \else $M_{SUSY}$\fi}}
\newcommand{\msusys}{{\ifmmode M_{SUSY}^2  \else $M_{SUSY}^2$\fi}}
\newcommand{\su}{{\ifmmode SU(3)_C\otimes\- SU(2)_L\otimes\- U(1)_Y \else $SU(3)_C\otimes SU(2)_L\otimes U(1)_Y$\fi}}
\newcommand{\suthree}{{\ifmmode SU(3)_C  \else $SU(3)_C$\fi}}
\newcommand{\sutwo}{{\ifmmode  SU(2)_L\otimes U(1)_Y \else $SU(2)_L\otimes U(1)_Y$\fi}}
\newcommand{\taup} {{\ifmmode \tau_{proton} \else $\tau_{proton}$\fi}}
\newcommand{\as}{{\ifmmode \alpha_{s}  \else $\alpha_{s}$\fi}}
\newcommand{\mgut}{{\ifmmode M_{GUT}  \else $M_{GUT}$\fi}}
\newcommand{\mguts}{{\ifmmode M_{GUT}^2  \else $M_{GUT}^2$\fi}}
\newcommand{\mze} {{\ifmmode m_0        \else $m_0$\fi}}
\newcommand{\mha}{{\ifmmode m_{1/2}    \else $m_{1/2}$\fi}}
\newcommand{\mb} {{\ifmmode m_{b}    \else $m_{b}$\fi}}
\newcommand{\mt} {{\ifmmode m_{t}    \else $m_{t}$\fi}}
\newcommand{\mts} {{\ifmmode m_{t}^2    \else $m_{t}^2$\fi}}

\newcommand{\mtau}{{\ifmmode m_{\tau}  \else $m_{\tau}$\fi}}
\newcommand{\dpp}{{\ifmmode \delta_{pert} \else $\delta_{pert}$\fi}}
\newcommand{\dnp}{{\ifmmode\delta_{non-pert}\else$\delta_{non-pert}$\fi}}
\newcommand{\dew}{{\ifmmode \delta_{\rm EW}\else $\delta_{\rm EW}$\fi}}
\newcommand{\rt}{{\ifmmode R_{\tau}  \else $R_{\tau} $\fi}}
\newcommand{\rz}{{\ifmmode R_{Z}  \else $R_{Z} $\fi}}

\newcommand{\swb}{{\ifmmode \sin^2\theta_{\overline{MS}} \else $\sin^2\theta_{\overline{MS}}$\fi}}
\newcommand{\cwb}{{\ifmmode \cos^2\theta_{\overline{MS}} \else $\cos^2\theta_{\overline{MS}}$\fi}}

\newcommand{\bsmm}{\ensuremath{B^0_s\to\mu^+\mu^-}}
\newcommand{\Bbsmm}{\ensuremath{{\mathcal Br}(\bsmm)}}

\newc\AIPCP[3] {{\em AIP Conf. Proc.} {\bf #1} (#2) #3}
\newc\AJ[3] {{\em Astrophys. J.} {\bf #1} (#2) #3}
\newc\AMS[3] {{\em Ann. Math. Statist.} {\bf #1} (#2) #3}
\newc\AP[3] {{\em Ann. Phys.} {\bf #1} (#2) #3}
\newc\APJ[3] {{\em Astropart. J.} {\bf #1} (#2) #3}
\newc\APP[3] {{\em Astropart. Phys.} {\bf #1} (#2) #3}
\newc\APS[3] {{\em Astrophys. J. Suppl.} {\bf #1} (#2) #3}
\newc\ARNPS[3] {{\em Ann. Rev. Nucl. Part. Sci.} {\bf C#1} (#2) #3}
\newc\BA[3] {{\em Bayesian Anal.} {\bf C#1} (#2) #3}
\newc\CPC[3] {{\em Comput. Phys. Commun.} {\bf C#1} (#2) #3}
\newc\CP[3] {{\em Contemp. Phys.} {\bf #1} (#2) #3}
\newc\EPJ[3] {{\em Euro. Phys. Journ.} {\bf C#1} (#2) #3}
\newc\JCAP[3] {{\em JCAP} {\bf #1} (#2) #3}
\newc\JHEP[3] {{\em JHEP} {\bf #1} (#2) #3}
\newc\JPG[3] {{\em J. Phys.} {\bf G #1} (#2) #3}
\newc\IJMP[3] {{\em Int. J. Mod. Phys.} {\bf A #1} (#2) #3}
\newc\MNRAS[3] {{\em Mon. Not. Roy. Astron. Soc.} {\bf #1} (#2) #3}
\newc\MPL[3] {{\em Mod. Phys. Lett.} {\bf A #1} (#2) #3}
\newc\NAR[3] {{\em New Astron. Rev.} {\bf #1} (#2) #3}
\newc\NCA[3] {{\em Nuovo Cimento} {\bf #1} (#2) #3}
\newc\NIM[3] {{\em Nucl. Instrum. Methods} {\bf #1} (#2) #3}
\newc\NIMA[3] {{\em Nucl. Instrum. Methods} {\bf A #1} (#2) #3}
\newc\NAT[3] {{\em Nature} {\bf #1} (#2) #3}
\newc\NPB[3] {{\em Nucl. Phys.} {\bf B #1} (#2) #3}
\newc\NPA[3] {{\em Nucl. Phys.} {\bf A #1} (#2) #3}
\newc\NPPS[3] {{\em Nucl. Phys. Proc. Suppl.} {\bf #1} (#2) #3}
\newc\PLB[3] {{\em Phys. Lett.} {\bf B #1} (#2) #3}
\newc\PR[3] {{\em Phys. Rep.} {\bf #1} (#2) #3}
\newc\PRL[3] {{\em Phys. Rev. Lett.} {\bf #1} (#2) #3}
\newc\PRD[3] {{\em Phys. Rev.} {\bf D #1} (#2) #3}
\newc\PRC[3] {{\em Phys. Rev.} {\bf C #1} (#2) #3}
\newc\PTP[3] {{\em Prog. Theor. Phys.} {\bf #1} (#2) #3}
\newc\RMP[3] {{\em Rev. Mod. Phys.} {\bf #1} (#2) #3 }
\newc\RPP[3] {{\em Rept. Prog. Phys.} {\bf #1} (#2) #3 }
\newc\SC[3] {{\em Science} {\bf #1} (#2) #3 }
\newc\ZPC[3] {{\em Z. Phys.} {\bf C #1} (#2) #3}
\newc\Err[3] {{\em Erratum-ibid.} {\bf #1} (#2) #3 }

\begin{document}
%\begin{frontmatter}
\begin{center}

\Large \textbf{ Constraints from the decay $\bsmm$ and LHC limits  on Supersymmetry}

\vspace{10mm}

\large

C. Beskidt$^1$, W. de Boer$^{1}$, D.I. Kazakov$^{2,3}$, F. Ratnikov$^1$, E. Ziebarth$^1$, V. Zhukov$^{1}$

\normalsize
\vspace{5mm}
$^1$ \textit{Institut f\"ur Experimentelle Kernphysik,
Karlsruhe Institute of Technology,\\ P.O. Box 6980, 76128 Karlsruhe, Germany}

\vspace{5mm}
$^2$ \textit{Bogoliubov Laboratory of Theoretical Physics, Joint Institute for Nuclear Research,\\
141980, 6 Joliot-Curie, Dubna, Moscow Region, Russia}

\vspace{5mm}
$^3$ \textit{Institute for Theoretical and Experimental Physics,\\
117218, 25 B.Cheremushkinskaya, Moscow, Russia}

\vspace{30mm} \textbf{Abstract} \vspace{5mm}

\begin{minipage}[c]{12cm}

\textit{
The pure leptonic decay  \bsmm\  is strongly suppressed in the Standard Model (SM), but can have large enhancements in Supersymmetry, especially at large values of \tb. New limits  on this decay channel from recent LHC data have been used to claim that these limits restrict the SUSY parameter space even more than the direct searches. However, direct searches are hardly dependent on \tb, while \Bbsmm\ is proportional to $\tan^6\beta$. The relic density constraint requires large {\tb} in a large region of the parameter space, which can lead to large values of {\bsmm}. Nevertheless, the experimental upper limit on \Bbsmm\ is not constraining the parameter space of the CMSSM more than the direct searches and the present Higgs limits, if combined with the relic density. We also observe SUSY parameter regions with negative interferences, where the \bsmm\ value is up to a factor three below the SM expectation, even at large values of {\tb}.
}
\end{minipage}
\end{center}

\thispagestyle{empty}
\setcounter{page}{0}

\section{Introduction}
Flavour Changing Neutral Currents (FCNC), like the leptonic decays of neutral B-mesons, are strongly suppressed in the Standard Model (SM), since they can only occur via loops involving the weak bosons. These decays are helicity suppressed, so the amplitudes are proportional to the mass of final state particles and the highest rates will be into tau leptons. The experimental signature for leptonic decays  is  clear:  search for an invariant mass in the mass window of the B-meson. This is easier for muonic decays. Hence, muonic B-decays have been investigated in much more detail at hadron colliders, especially since these decays can be strongly enhanced by loop corrections involving particles beyond the SM, like Supersymmetry \cite{Choudhury:1998ze,Huang:1998vb,Babu:1999hn,Dedes:2001fv,Bobeth:2002ch,Arnowitt:2002cq}. The \bsmm\ decay mode
 has received significant attention \cite{Dutta:2011bk,Akula:2011ke,Hooper:2011tk} after the CDF collaboration announced a measurement a factor five to six above the expected SM value \cite{Aaltonen:2011fi}. However, the excess was not confirmed by subsequent LHC measurements \cite{cmslhcb}, but nevertheless the LHC upper limit can give significant constraints on
the SUSY parameter space, see e.g. \cite{Akeroyd:2011kd}, where in some scenarios better limits than those obtained from direct searches have been claimed. However, the excluded parameter space depends strongly on the choice of \tb\, since the \bsmm\ rate varies as $\tan^6\beta$. The relic density constraint correlates \tb\ with the SUSY mass parameters \cite{Beskidt:2010va}, so if one combines the cosmological constraint with the accelerator constraints there is no arbitrary choice for \tb\ anymore.
Although the relic density requires a large value of \tb\ in a large region of parameter space, we show that the excluded SUSY mass ranges are well below the LHC constraints from direct searches \cite{Aad:2011xm,Aad:2011hh,Collaboration:2011zy,Chatrchyan:2011qs} and the limits on the pseudo-scalar Higgs \cite{Aad:2011rv,cmsma}. In principle, other constraints, like g-2, {$b \to s \gamma$} and {$B \to \tau \nu$} could also be considered. However taking these into account requires a careful treatment of the non-gaussian systematic errors. which is beyond the scope of the present letter. Numerous studies combining these variables with the recent LHC data have appeared \cite{Buchmueller:2011ki,Bertone:2011nj,Sekmen:2011cz}. 
%An interesting example is shown in the two top diagrams of Fig. \ref{f7}. 
\begin{figure}[]
\begin{center}
  \includegraphics[width=0.7\textwidth]{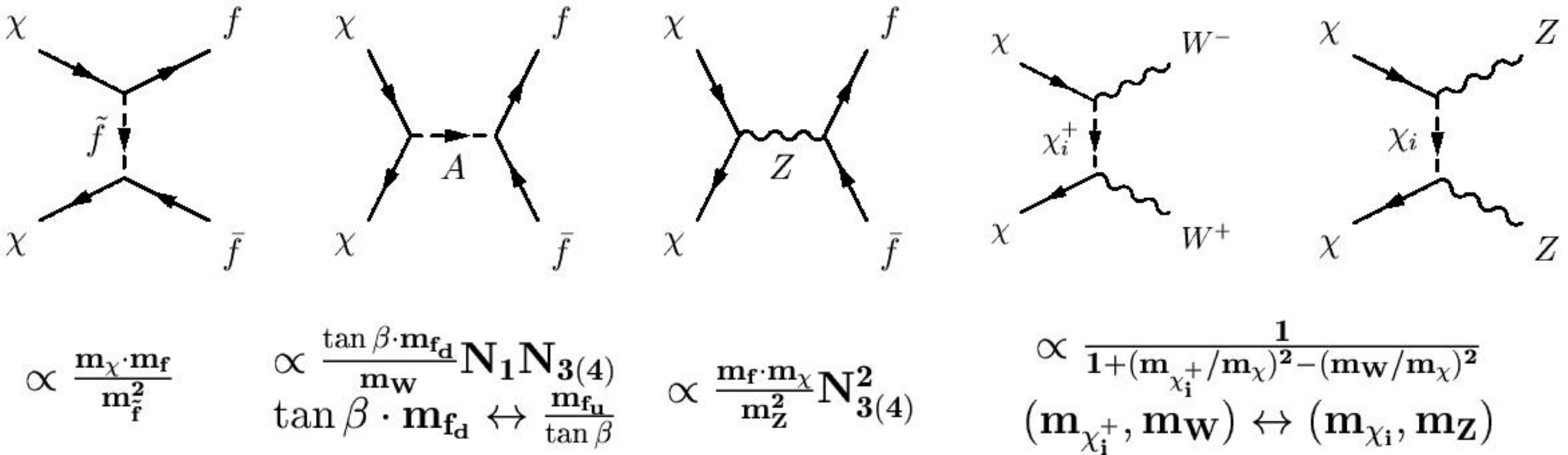}
\caption[]{ Annihilation diagrams for  the lightest neutralino, which is
  a linear combination of the gaugino and Higgsino states:
  $|\chi_o>=N_1 |B_0>+N_2|W^3_0>+N_3|H_1>+N_4|H_2>$. The dependence of the amplitudes on  masses
  and neutralino mixing parameter $N_i$ has been indicated. }
\label{f1}
\end{center}
\end{figure}

\section{Relic Density}\label{relic} 
  The relic density and annihilation cross section $\sigma$ are related through:
\begin{equation}
\Omega h^2=\frac{3.10^{-27}}{<\sigma v>},
\label{e1}
\end{equation}
where the annihilation cross section $\sigma$ averaged over the relative velocities of the neutralinos is given in pb \cite{Kolb:1990vq,Jungman:1995df} and $h\approx 0.71$ is the Hubble constant in units of 100 $(km/s)/Mpc$.  The best value for the relic density is $\Omega h^2=0.1131\pm 0.0034$ \cite{Komatsu:2010fb}. 
%For the relation between cross section and expansion rate in Eq. \ref{e1} one assumes DM was a thermal relic, which froze out at the time when the annihilation rate was about equal to the expansion rate, given by the Hubble constant.
For a given relic density $\Omega$ the annihilation cross section is known
independent of a specific model, since it only depends on the observed Hubble constant and the observed relic density. Its value is furthermore largely independent of the neutralino mass $m_\chi$ (except for logarithmic corrections)\cite{Kolb:1990vq,Jungman:1995df}.
The DM constraint should exist for any model, but to be specific  the
 Constrained Minimal Supersymmetric Standard Model (CMSSM) with supergravity
inspired breaking terms, will be considered \cite{Chamseddine:1982ah,Nath:1982zq,Hall:1983iz}. It is characterized by  5 parameters:
$m_0,~m_{1/2},~\tb,~\mbox{sign}(\mu), ~A_0$. Here $m_0$ and $m_{1/2}$ are
the common masses for the gauginos and scalars at the GUT scale, which is determined by
the unification of the gauge couplings at this scale. Gauge unification is perfectly possible with the
latest measured couplings at LEP~\cite{deBoer:2003xm}.
%The ratio of the vacuum expectation values
%of the neutral components of the two Higgs doublets in Supersymmetry is called $\tb$ ~ and
%$A_0$ is the trilinear coupling at the GUT scale.
Electroweak symmetry breaking (EWSB) fixes the scale of $\mu$
\cite{Inoue:1982pi}, so only its sign is a free parameter.
  The positive sign is taken, as
suggested by the small deviation of the SM prediction from the muon anomalous moment.

The relic density can be calculated from the diagrams in Fig.   \ref{f1}.
For its calculation we used the public code micrOMEGAs 2.4 \cite{Belanger:2010pz,Pukhov:2010px} combined with Suspect 2.41 as  mass spectrum calculator \cite{Djouadi:2002ze}. The optimal parameters were found by minimizing the $\chi^2$ function using the Minuit program \cite{James:1975dr}.

For heavy SUSY masses the sfermion exchange diagram is suppressed, 
 the W- and Z-final states from t-channel chargino and
neutralino exchange have  a small cross section, the coupling of the LSP to the Z-boson is only via the Higgs component of the LSP, which is typically small, so in most regions of  parameter space  the  pseudo-scalar
Higgs exchange is dominant, except for the co-annihilation regions. These are the regions, where the Next-to-Lightest Supersymmetric Particle (NLSP) and LSP are nearly mass-degenerate. In this case they co-exist in the early universe until the common freeze-out temperature and can co-annihilate. This happens if the stau and neutralino are degenerate and co-annihilate into a tau \cite{Ellis:1999mm}. For this to happen the values of $m_0$ and $m_{1/2}$ have to be fine-tuned to a high degree, so it happens only in a thin stripe in the $m_0-m_{1/2}$ plane, as will be shown below. Another co-annihilation region happens at the border of parameter space, where electroweak symmetry breaking does not occur anymore, since here the Higgs mixing parameter becomes negative. In the transition region $\mu$ becomes small and the lightest chargino and lightest neutralino become nearly degenerate Higgsinos, as is obvious from the mass matrices, which have as lowest eigenvalues either a gaugino mass term or a Higgsino mass term, if the mixing is neglected. In this case gauginos can co-annihilate into a W-boson \cite{Griest:1990kh}. 

\begin{figure}
 \begin{center}
  \includegraphics[width=0.45\textwidth]{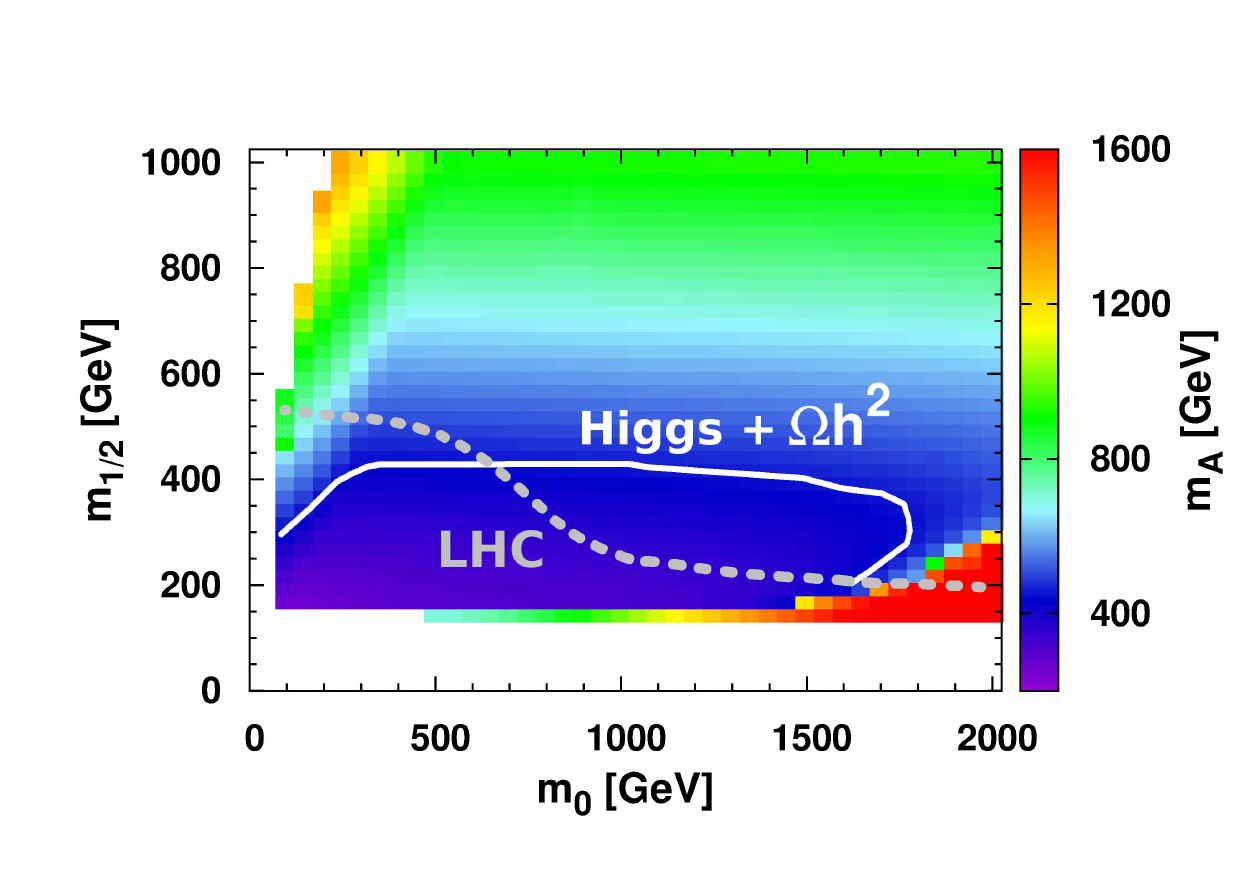}
     \includegraphics[width=0.45\textwidth]{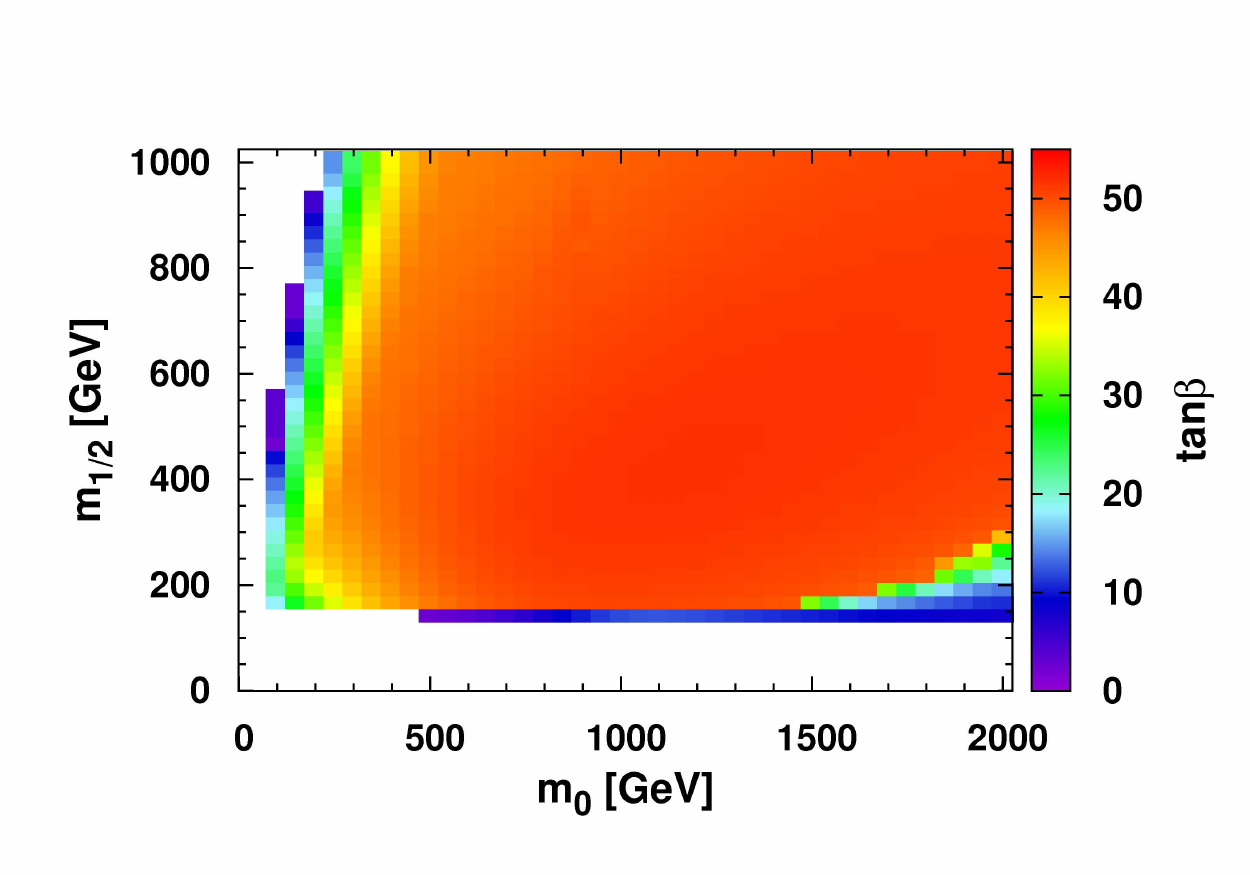}
 \end{center}%\vspace{-0.2cm}
 \caption{The value of the pseudo-scalar Higgs mass (left) and the value of  $\tan\beta$ required for a correct relic density
 in the  $m_0-m_{1/2}$ plane. The excluded region by the limit on the pseudoscalar Higgs mass from Ref. \cite{cmsma}  is indicated by the white solid  line in the left panel, while the dashed line indicates the limit from the direct searches from Ref. \cite{Collaboration:2011zy}. The \tb\ value is around 50 in the central region, as indicated by the colour coding in the right panel and decreases towards the edges, where co-annihilation starts to be important. }\label{f2}
 \end{figure}

Outside the bulk region with low SUSY masses and the co-annihilation regions the dominant contribution comes from A-boson exchange: $\chi+\chi \to A \to b\bar b$,
 which  is proportional to
\beq <\sigma v>\sim  \frac{m_\chi^4
m_b^2 \tan^2\beta}{\sin^4 2\theta_W
\,M_Z^2}\frac{ \left( N_{31}\sin\beta -N_{41}\cos\beta
\right)^2\left( N_{21}\cos\theta_W - N_{11}\sin\theta_W
\right)^2}{\left( 4m_\chi^2 - m_A^2
\right)^2+m_A^2\Gamma_A^2}.\label{eq2}
\eeq
The elements of the mixing matrix in the neutralino sector define the content of the lightest neutralino:
$$
|\tilde \chi^0_1\rangle =N_{11}|B_0\rangle
+N_{21}|W^3_0\rangle +N_{31}|H_1\rangle +N_{41}|H_2\rangle .$$
The sum of the diagrams should yield $<\sigma v>=2\cdot  10^{-26}\ cm^3/s$ to get the correct relic density, which implies that the annihilation cross section $\sigma$ is of the order of a few pb. Such a high cross section can be obtained only close to the resonance, i.e. $m_A\approx 2m_\chi$. Actually on the resonance the cross section is too high, so one needs to be in the tail of the resonance, i.e. $m_A\approx 2.2m_\chi$ or $m_A\approx 1.8 m_\chi$. So one expects $m_A \propto m_{1/2}$ from the relic density constraint. This is shown in the left panel of Fig. \ref{f2}. Here we optimized simply \tb\ for each pair of $m_0-m_{1/2}$ values, as was done in Ref. \cite{Beskidt:2010va}. The corresponding values of \tb\ needed are shown in the right panel of Fig. \ref{f2}. The production cross section of the pseudo-scalar Higgs at the LHC is proportional to $\tan^2\beta$, so the present limits from LHC are proportional to  $\tb^2$ as well. At \tb =50 the present limit on $m_A$ is about 450 GeV, so the limit on $m_{1/2}$ is close to it. The exclusion on $m_A$  as function of \tb\ from the CMS collaboration \cite{cmsma} is
indicated in Fig. \ref{f2} as well. Similar limits have been obtained by the ATLAS collaboration \cite{Aad:2011rv}.  Also the excluded region from the direct searches has been indicated
using the CMS data \cite{Collaboration:2011zy}. Similar results were obtained by other searches \cite{Aad:2011xm,Aad:2011hh,Chatrchyan:2011qs}.
 \begin{figure}%[tbh]
 \begin{center}
  \includegraphics[width=0.45\textwidth,height=0.3\textwidth]{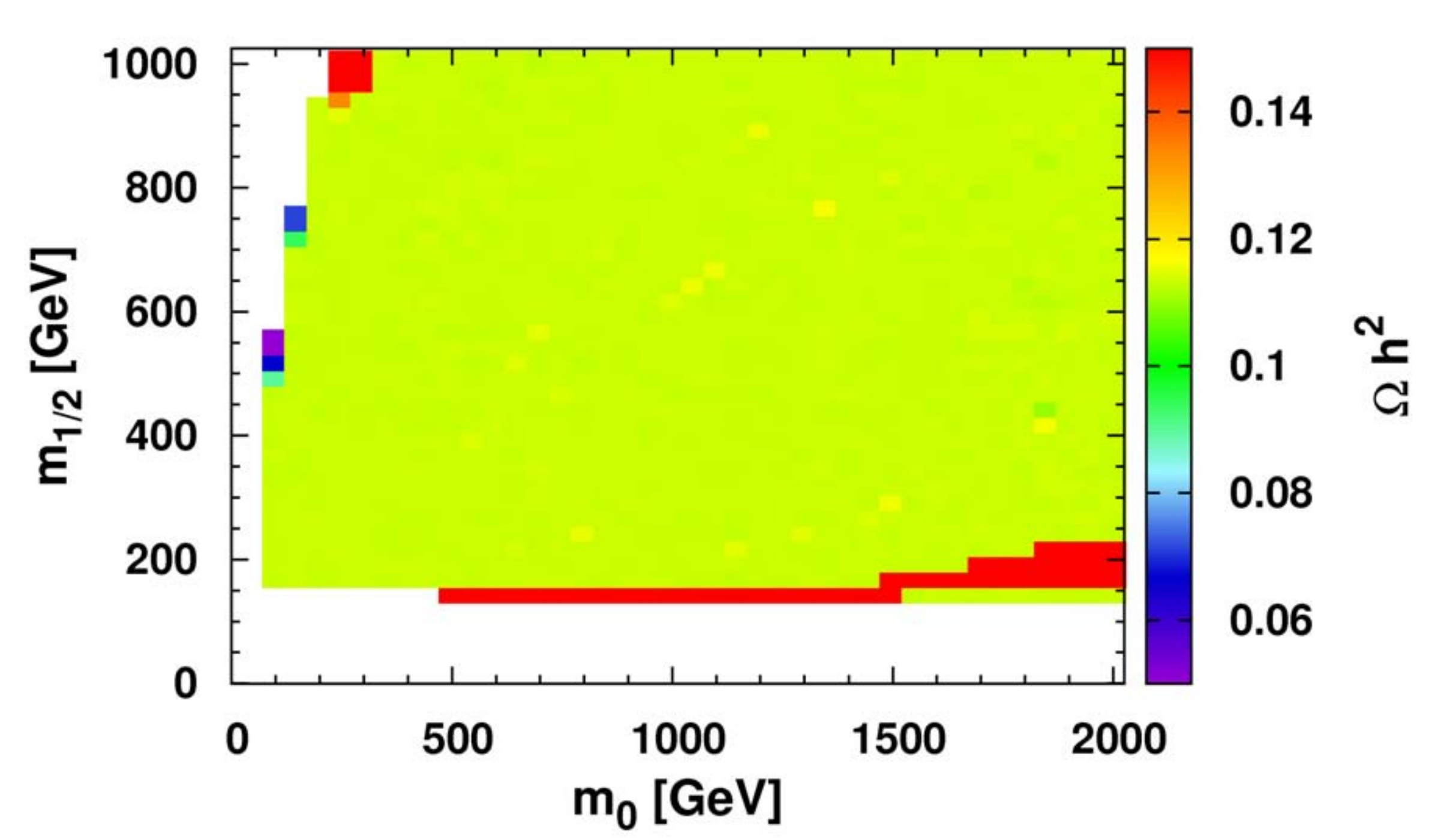}
     \includegraphics[width=0.45\textwidth,height=0.3\textwidth]{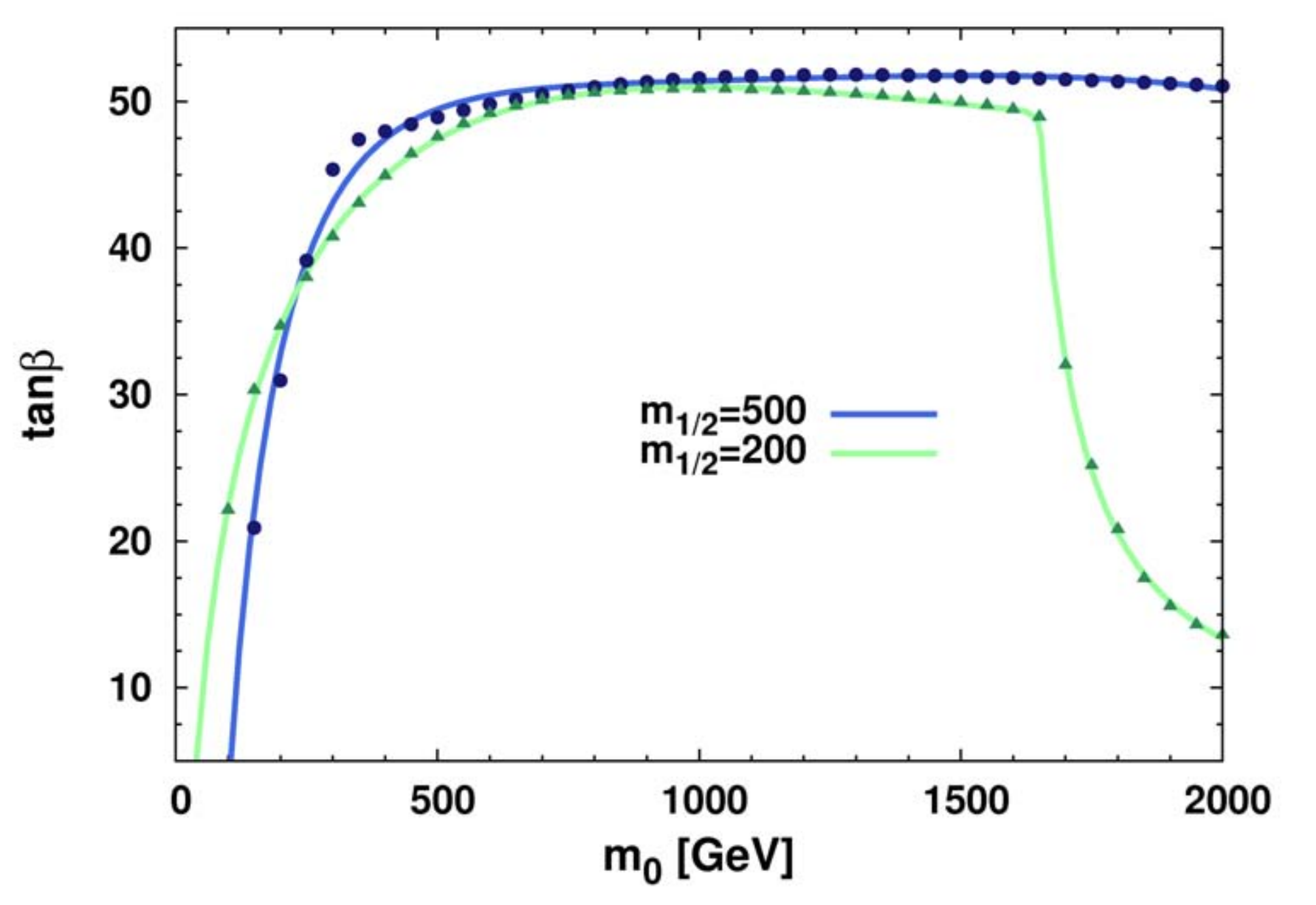}
 \end{center}%\vspace{-0.2cm}
 \caption{The relic density in the $m_0-m_{1/2}$ plane (left)  
 and $\tan\beta$ as function $m_0$ for different values of $m_{1/2}$ (right) The  colour coding in the left panel shows that the relic density constraint can be fulfilled for all SUSY masses.
  }\label{f3}
 \end{figure}
The relic density constraint can be fulfilled with the parameters of Fig. \ref{f2}, as demonstrated 
in the left panel of Fig. \ref{f3}. The top left is excluded, since here the LSP is not a neutral particle, but the stau is the LSP.
%The bottom right is excluded, since here EWSB does not work, because  $\mu^2$ becomes negative, where $\mu$ is the mixing parameter in the Higgs potential. In the transition to these forbidden regions
%co-annihilation of the LSP with an almost mass-degenerate other SUSY particle is possible. 
%The co-annihilation and other properties of these transition regions have been reviewed nicely in Ref. \cite{nanopoulos} and the reader is referred to this reference for details and original references. 
In the co-annihilation regions the annihilation via the pseudo-scalar Higgs exchange has to be suppressed, thus requiring a larger value of $m_A$ and a corresponding lower value of \tb, as demonstrated in the right panel of Fig. \ref{f3}. It is just a more detailed plot of the right hand panel of Fig. \ref{f2} for two values of $m_{1/2}$.

In summary, if one allows $\tb$ to vary in the $m_0-m_{1/2}$ plane, one obtains the observed relic density for $any$ combination of $m_0$ and $m_{1/2}$, i.e. the relic density allows $all$ masses for the SUSY   sparticles.
However, the \bsmm\ constraint has to be investigated for the large values of  \tb\ required by the relic density.

\section{\bsmm\ decay rate}\label{bsmumu}
\begin{figure}[]%\vspace{-0.4cm}
\begin{center}
   \includegraphics[width=0.32\textwidth,height=0.2\textwidth]{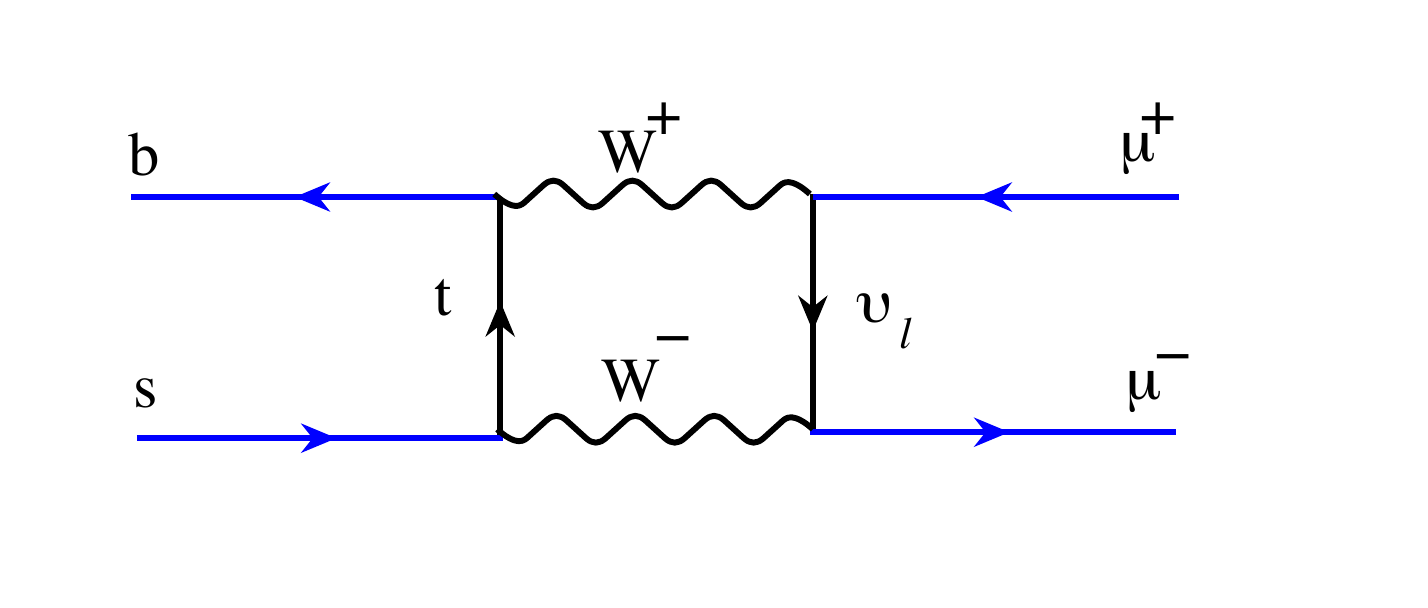}
    \includegraphics[width=0.32\textwidth,height=0.2\textwidth]{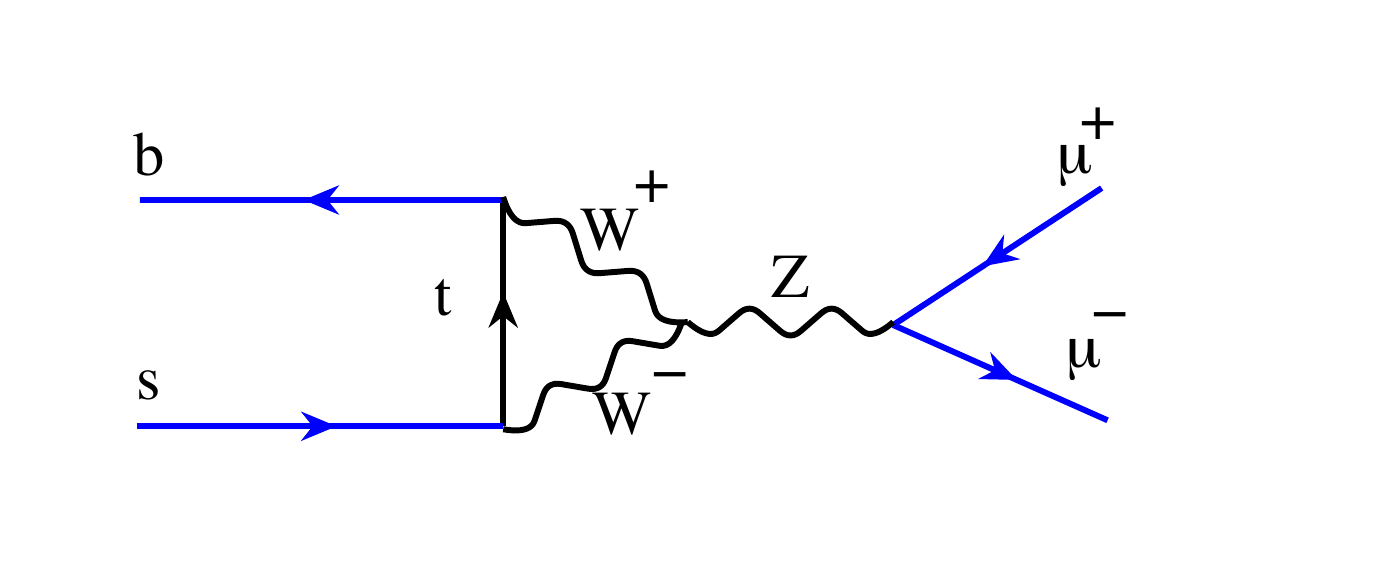}
   \includegraphics[width=0.32\textwidth,height=0.2\textwidth]{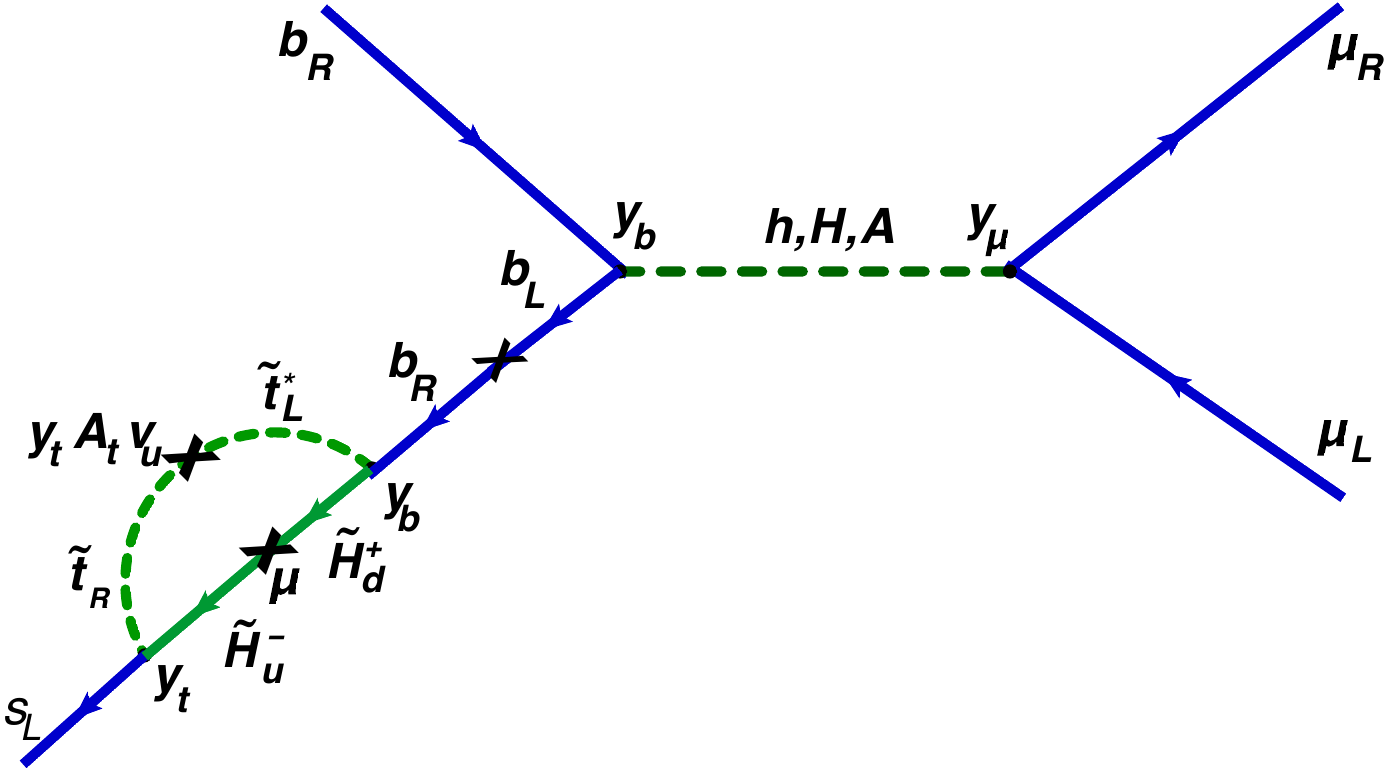}
\end{center}
\caption{The diagrams contributing to the \bsmm\ decay in the SM and in the MSSM.} \label{f7}
\end{figure}

% In the SM the  decay $B_s\to \mu^+\mu^-$ is given by the diagrams shown on top of  Fig.\ref{f7}. The branching ratio is \cite{Anik} $${\mathcal BR}^{SM}(B_s\to\mu^+\mu^-)=3.5\cdot  10^{-9}$$, while  experiments only provide an upper bound\cite{bmu} $${\mathcal BR}^{Ex}(B_s\to\mu^+\mu^-)<4.5 \cdot 10^{-8}.$$  In the MSSM one has several diagrams, but the main SUSY  enhancement  at large $(\tan\beta)$ $(\propto \tan\beta)^6$~(!) comes from the one  shown at the bottom of  Fig.\ref{f7}.

\begin{figure}[]%\vspace{-0.4cm}
\begin{center}
   \includegraphics[width=0.45\textwidth]{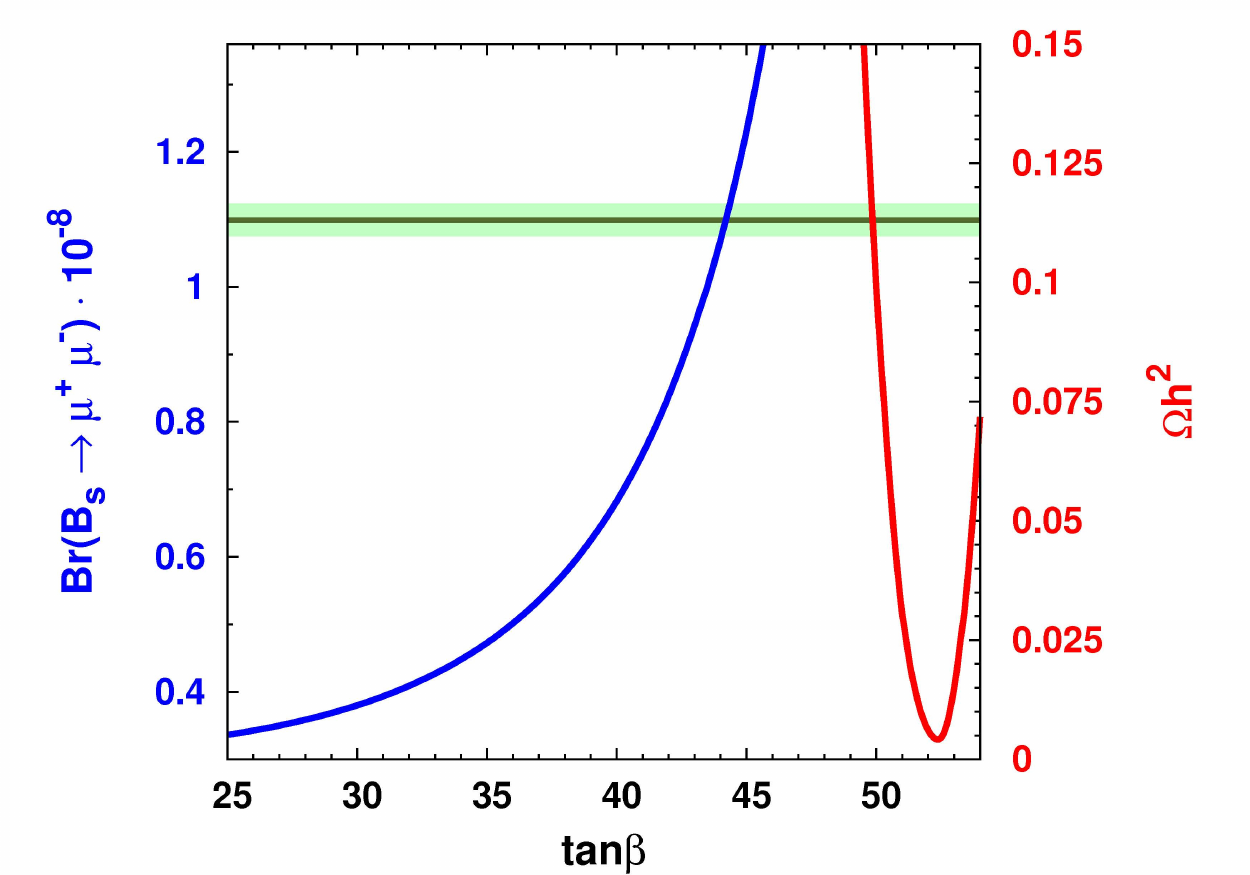}
   \includegraphics[width=0.45\textwidth]{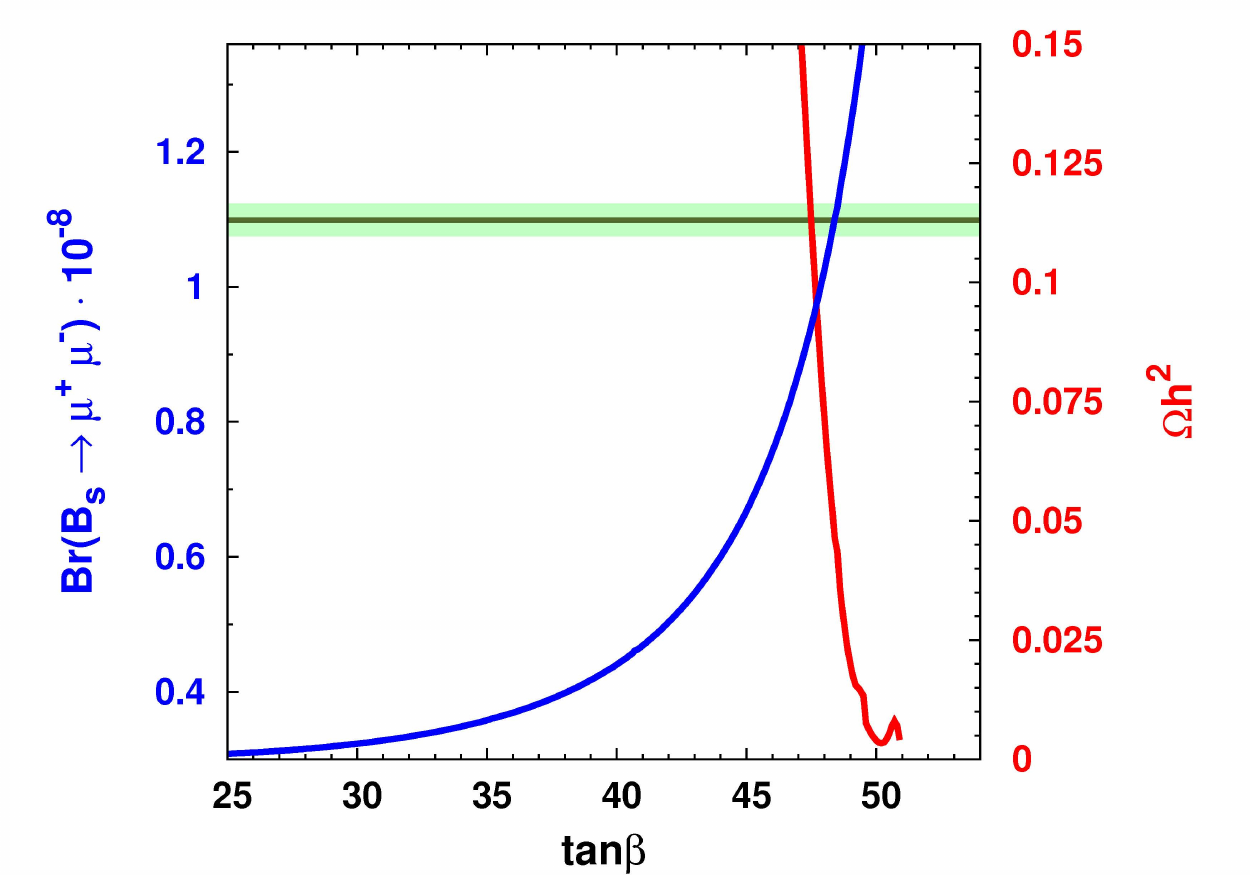}
\end{center}
\caption{The \tb\ dependence of \Bbsmm\ and the relic density for $A_0$=0 (left) and $A_0>0$ (right).
The left and right vertical scales are for  \Bbsmm\ and  $\Omega h^2$, respectively
and the scales have been adjusted so, that the horizontal line indicates the upper limit for \Bbsmm\ and the observed value for $\Omega h^2$.} \label{f8}
\end{figure}

  The branching ratio for $\bsmm$ is taken from Ref.
\cite{Arnowitt:2002cq}, which we write in the form
\beqa  \Bbsmm  &=&   {{2\tau_B m_B^5}\over{64\pi}}f^2_{B_s}\sqrt{1-{{4m_l^2}\over{m_B^2}}}
	\nonumber \\
	& &
 \left[ \left( 1-{{4m_l^2}\over{m_B^2}} \right)
	\left| {{(C_S-C_S')}\over{(m_b+m_s)}} \right|^2+
 \left| {{(C_P-C_P')}\over{(m_b+m_s)}}+
	2{m_{\mu}\over m_{B_s}^2}(C_A-C_A') \right|^2
	\right]~~
\label{eq:bsmm_br_msugra}
\eeqa
where $f_{B_s}$  is the $B_s$ decay constant,
$m_{B}$ is
the $B$ meson mass, $\tau_B$ is the mean life and $m_l$ is the mass of lepton. 
$C_A$, $C^{\prime}_A$ are largely determined by the SM diagrams, while
$C_S$, $C^{\prime}_S$, $C_P$,
$C^{\prime}_P$ include the SUSY loop contributions  due to diagrams involving
particles such as stop, chargino, sneutrino, Higgs etc..  
%For large $\tan\beta$, the  amplitude has terms that grow like $\tan^{3}\beta$.  
For large $\tan\beta$ values the dominant contribution to
$C_S$ can be written as:
\beq
 C_S \simeq
 {{G_F\alpha}\over {\sqrt 2\pi}}V_{tb}V_{ts}^*
 \left( {\tan^3\beta\over{4\sin^2\theta_W}} \right)
 \left({{m_bm_{\mu} m_t \mu}\over{M_W^2 M_A^2}} \right)
 {\sin2\theta_{\tilde t}\over 2} \left( \frac{ m_{\tilde t_1}^2
\log\left[\frac{m_{\tilde t_1}^2}{\mu^2}\right] }
	{ \mu^2-m_{\tilde t_1}^2 } -
  \frac{ m_{\tilde t_2}^2 \log\left[\frac{m_{\tilde t_2}^2}{\mu^2}\right] }
	{ \mu^2-m_{\tilde t_2}^2 } \right) \label{bmm}
\eeq
where $m_{\tilde t_{1,2}}$ are the two stop masses,  and
$\theta_{\tilde t}$  is the rotation angle  to diagonalize the stop mass matrix.
 We need to multiply  the above expression by
$1/(1+\epsilon_b)^2$ to include the SUSY QCD corrections, where $\epsilon_b$ is
proportional to  $\mu\tan\beta$ \cite{Carena:1999py}.  We have $C_P = -C_S,  C^{\prime}_S =
(m_s/m_b)C_s$  and $C^{\prime}_P = - (m_s/m_b)C_P$.
One observes from Eq. \ref{bmm} the $\tan^6\beta$ dependence, but one also observes the strong suppression in the last term if the stop masses become equal. 
In the MSSM the stop mass splitting is given by (see e.g. reviews \cite{deBoer:1994dg,Kazakov:2010qn}:
\begin{equation}
\tilde{m}^2_{1,2}= \frac 12 \left(\tilde m_{tL}^2+\tilde m_{tR}^2
\pm \sqrt{ (\tilde m_{tL}^2-\tilde m_{tR}^2)^2+4m_t^2(A_t-\mu
\cot\beta})^2\right),
 \label{stop}
\end{equation}
where the left- and right-handed quark masses are defined by:
\begin{eqnarray*}
  \tilde m_{tL}^2&=&\tilde{m}_Q^2+m_t^2+\frac{1}{6}(4M_W^2-M_Z^2)\cos 2\beta ,\\
  \tilde m_{tR}^2&=&\tilde{m}_U^2+m_t^2-\frac{2}{3}(M_W^2-M_Z^2)\cos 2\beta .
\end{eqnarray*}
For large SUSY scales the mass terms for the right-handed singlet $m_U$ and left-handed doublet $m_Q$ become large and $m_{tL}$ and $m_{tR}$ become of the same order of magnitude. Then the stop splitting is determined by the term $A_t-\mu/\tan\beta$, so for large $\tan\beta$ the second term is small and the stop mixing can be made small by increasing the trilinear coupling $A_0$ at the GUT scale. 
%which leads to negative values of $A_t$ at low energies by the  running. 
One indeed can eliminate the  tension between the large value of $\tan\beta$ required by $\Omega h^2$ and the \bsmm\ rate, as demonstrated in Fig.  \ref{f8}: in the left (right) panel the dependence of \Bbsmm\ and  $\Omega h^2$ are shown as function of \tb\ for $A_0=0$ ($A_0 >0$). The left and right vertical scales are for  \Bbsmm\ and  $\Omega h^2$, respectively
and the scales have been adjusted so, that the horizontal line indicates the upper limit for \Bbsmm\ and the observed value for $\Omega h^2$. One observes from the left panel that for the correct value of \tb =50 for $\Omega h^2$ the value of \Bbsmm\ is far above the experimental upper limit, but if one adjusts $A_0$ both can be brought into agreement (right panel). Here we fitted simply $A_0$ and {\tb} for each value of $m_0$ and $m_{1/2}$ in the $m_0 - m_{1/2}$ plane with \bsmm and $\Omega h^2$ as constraint.
\begin{figure}
 \begin{center}
 %\leavevmode
   \includegraphics[width=0.45\textwidth]{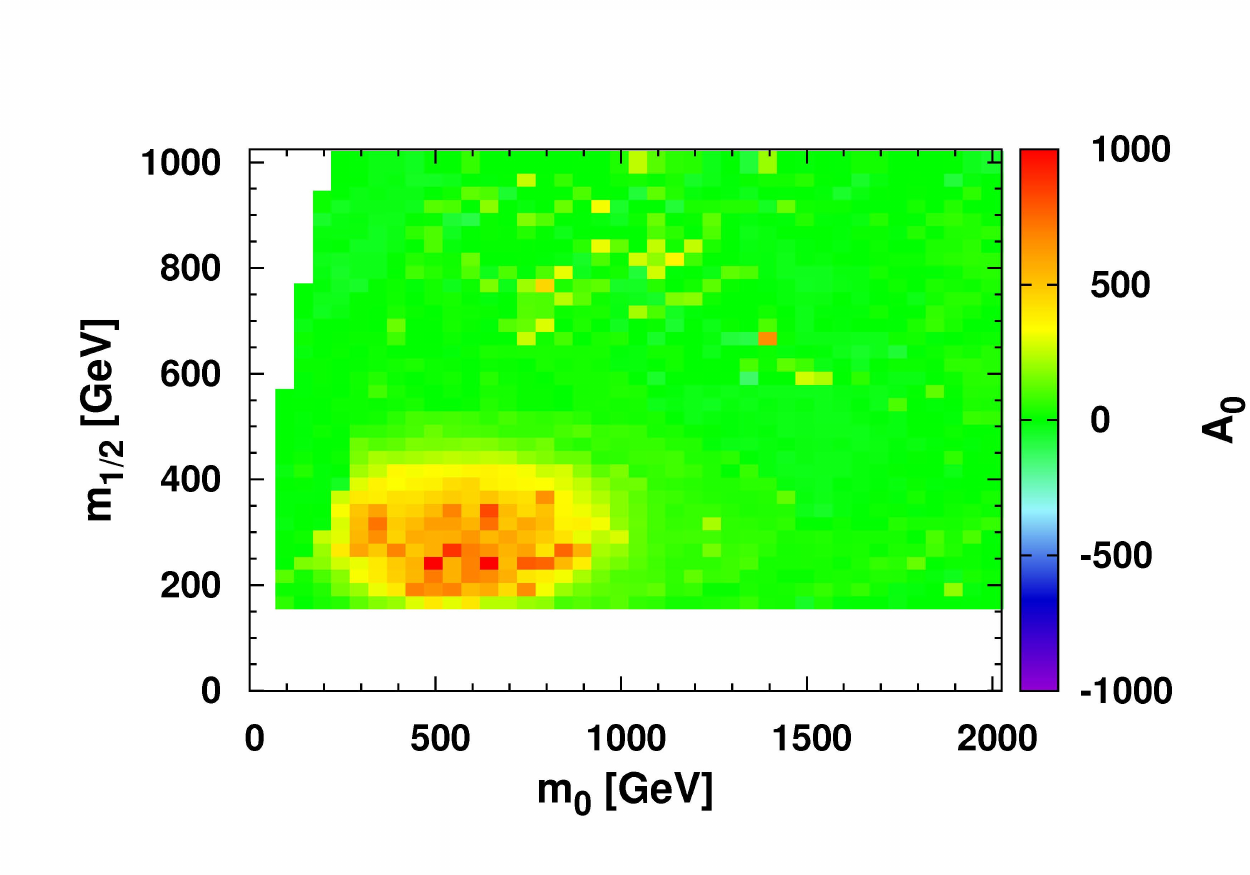}
    \includegraphics[width=0.45\textwidth]{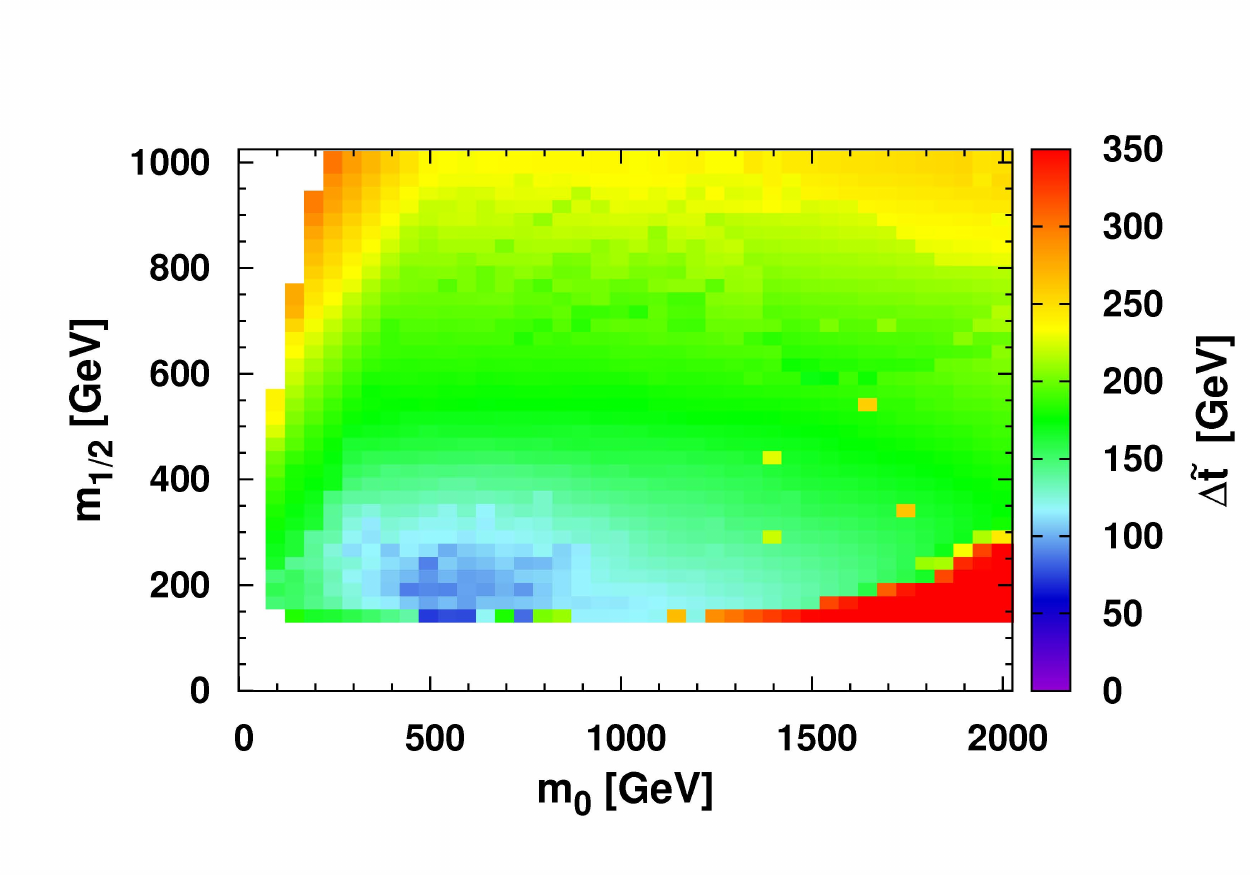}
 \end{center}\vspace{-0.3cm}
 \caption{The \Bbsmm\ constraint can lead to tension in combination with the relic density constraint, since the latter requires large $\tan\beta$, which leads to a large stop splitting. However, this can be compensated with a large value of $A_0$ (left panel), which reduces the difference between the stop masses $\Delta\tilde{t}$ (right panel) in the region where otherwise the constraint
$\Bbsmm<4.7 \cdot 10^{-8}$ could not be fulfilled.
}
 \label{f9}
 \end{figure}
 \begin{figure}
 \begin{center}
 %\leavevmode
   \includegraphics[width=0.45\textwidth]{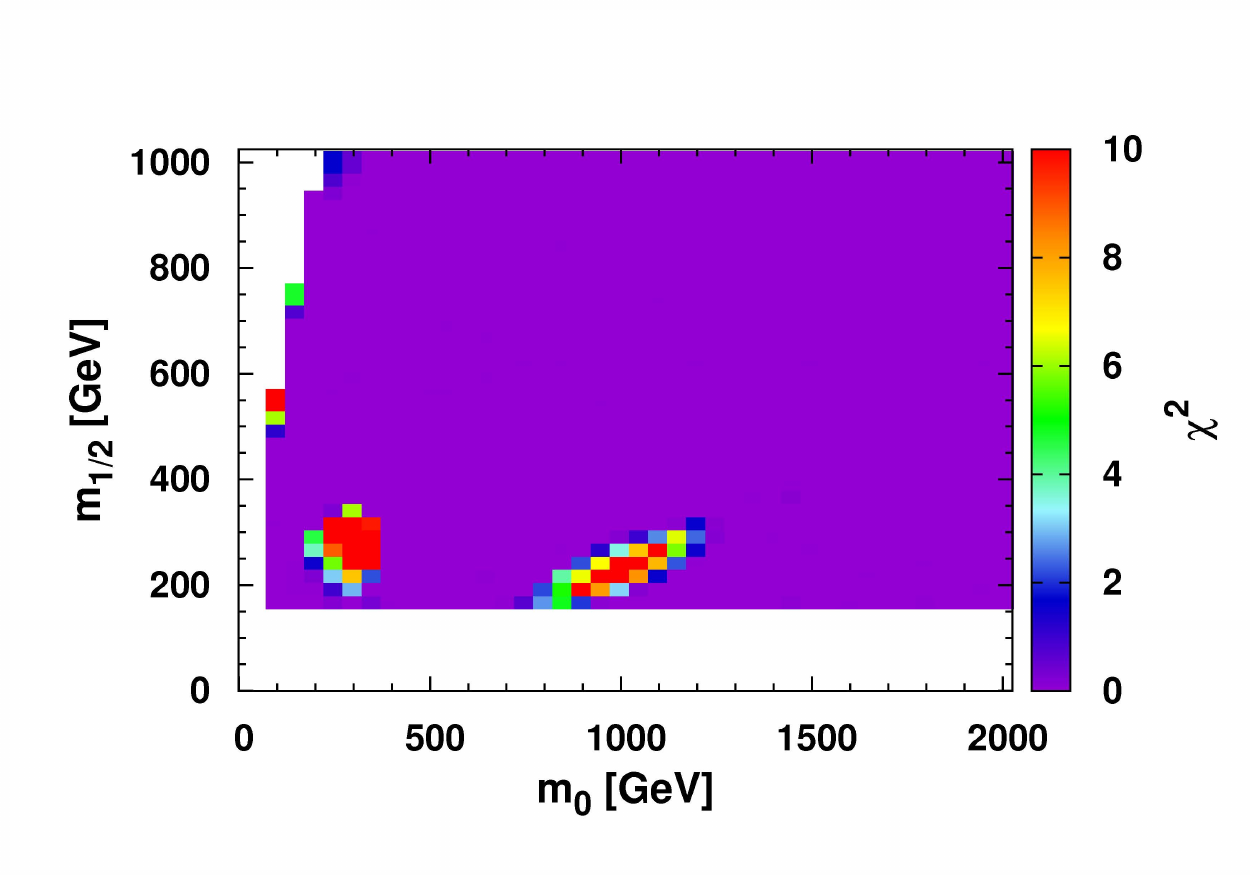}
    \includegraphics[width=0.45\textwidth]{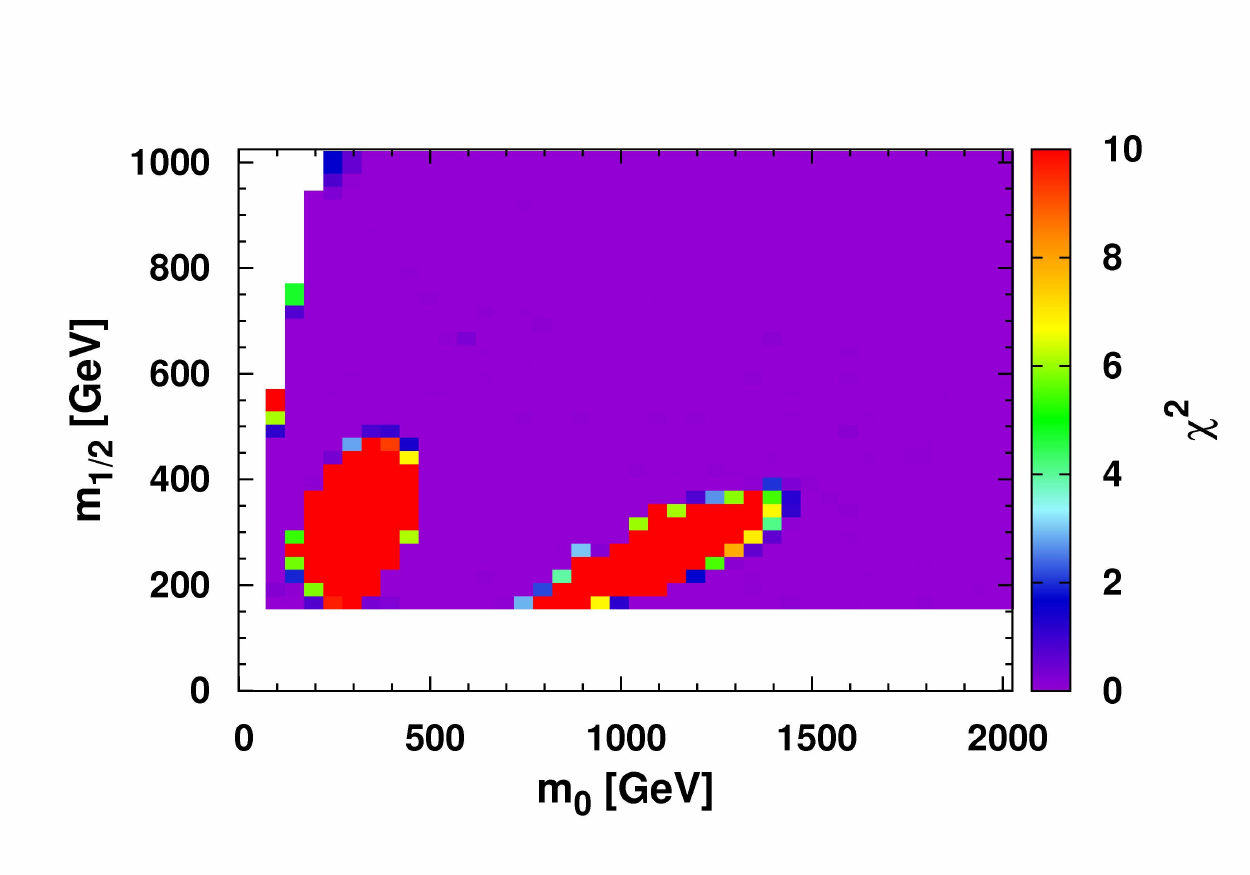}
 \end{center}\vspace{-0.3cm}
 \caption{Excluded region from a combined fit of the relic density and the upper limit on $\Bbsmm<1.1 \cdot 10^{-8}$ (left)
  and a hypothetical $\Bbsmm<0.66 \cdot 10^{-8}$ (right). The colour code indicates the $\chi^2$ value. $\chi^2$=5.99 indicates the 95\% C.L. contour, which essentially corresponds to the red region.
}
 \label{f10}
 \end{figure}
  \begin{figure}
 \begin{center}
 %\leavevmode
   \includegraphics[width=0.45\textwidth]{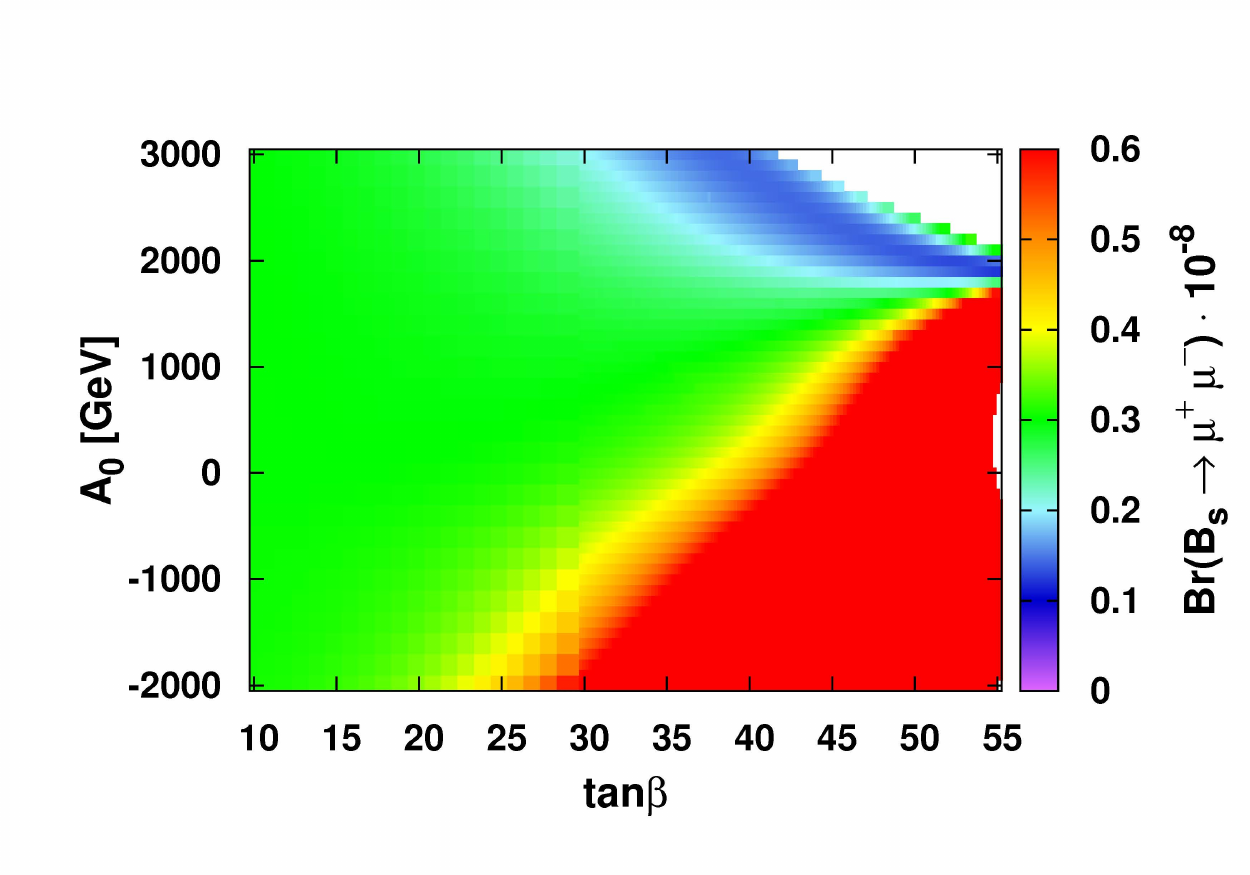}
    \includegraphics[width=0.45\textwidth]{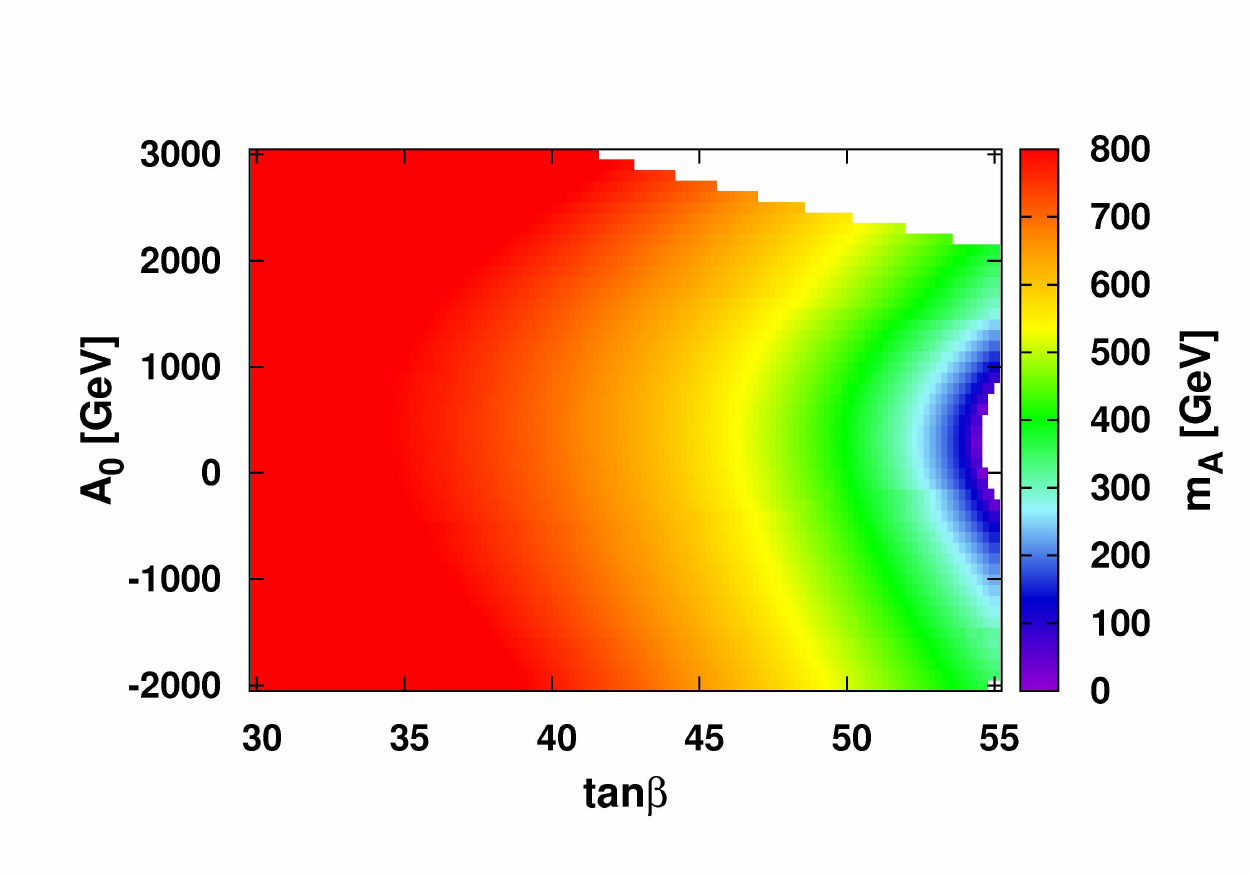}
 \end{center}\vspace{-0.3cm}
 \caption{\Bbsmm\ and the pseudoscalar Higgs mass $m_A$ as function of $A_0$ and \tb\ 
 in the left and right panel, respectively. The figures corresponds to $m_0=1000$ GeV and $m_{1/2}=250$ GeV, which is inside the excluded region on the right hand side of Fig. \ref{f10}. Note that the green region in the left panel corresponds to the SM value, while the blue (red) region corresponds to values below (above) the SM value.
}
 \label{f11}
 \end{figure}
The fitted values of $A_0$ reduce the stop mass to low enough values to force agreement. The required values of $A_0$ and the corresponding stop mass differences are shown in Fig. \ref{f9}.
The excluded regions in the combined fit of the relic density and \Bbsmm\ are shown in Fig. \ref{f10} for the present limit (left panel) and for a hypothetical limit of twice the SM value (right panel). One observes that the limit from \Bbsmm\ is well below the limits from the direct and Higgs searches shown in \mbox {Fig. \ref{f2}}.  The reason for the two-lobed  excluded regions is the following: at small values of $m_0$ the trilinear coupling cannot be made large enough to suppress \Bbsmm\ enough, because the staus become tachyonic. At intermediate values of $m_0$ the trilinear couplings can be made large enough, but at larger values of $m_0$ the pseudoscalar Higgs boson mass $m_A$ becomes too large for large $A_0$ values (see Fig. \ref{f11} right) and the relic density becomes too large as well. For values of $m_0$ well above 1 TeV the loop contributions are suppressed  enough to fulfill the \Bbsmm\ constraint.
These results are demonstrated in Fig. \ref{f11}, which displays the values of \Bbsmm\  and the pseudoscalar Higgs mass in the $A_0$-\tb\ plane for $m_0=1000$ and $m_{1/2}=250$ GeV, i.e. in the excluded lobe on the right hand side in Fig. \ref{f10}.
The green region in the left pannel of Fig. \ref{f11} corresponds to values close to the SM value for \Bbsmm,\ but at large positive values of $A_0$ and large values of \tb\ the \Bbsmm\ value drops {\it below} the SM value (blue upper right region), while at lower values of $A_0$ one observes the famous large \tb\ enhancement (red bottom right region). In the right top corner the staus become tachyonic, so this theoretically disfavored region is left white.
Surprisingly, values of \Bbsmm\ can fall up to a factor three {\it below} the SM value,
 which can be explained as follows. In Eq. \ref{bmm} $\sin(2\theta_{\tilde t})$ can change sign, depending on the value of the   off-diagonal element in the stop mixing matrix $A_t-\mu/\tb$. Hence, $C_P$ can change sign as well and the term  
  $\left| {{(C_P-C_P')}/{(m_b+m_s)}}+
	2{m_{\mu} /m_{B_s}^2}(C_A-C_A') \right|^2$  in Eq. \ref{eq:bsmm_br_msugra} can become small, if $C_P$ and $C_A$ have opposite sign.
We have checked that this change in sign is indeed the origin of the negative interference between the SM value and the SUSY values, both in the micrOMEGAs code, which we used, and in the SuperIso V3.1 code \cite{Arbey:2009gu}, which gives almost identical results.

\section{Summary}
We have calculated the excluded regions in the CMSSM from the recent upper limits on the \bsmm\ decays in combination with the relic density constraint. The latter requires large \tb\ values in the regions outside the co-annihilation regions and since \Bbsmm\ is proportional to $\tan^6\beta$ one could expect strong constraints from the recent upper limits. However, the \Bbsmm\ approaches zero in case the splitting between the stop1 and stop2 masses approaches zero. This splitting is determined by the off-diagonal element $A_t-\mu/\tan\beta$ of the stop mixing matrix, which can be made small for large \tb\ and a positive value of the trilinear coupling $A_0$ at the GUT scale. From a simultaneous fit of $A_0$ and \tb\
to the combined data of \Bbsmm\ and relic density we find the excluded regions from these constraints to be well below the constraints  from the Higgs searches and direct searches at the LHC. This holds even in the case that a hypothetical limit on \Bbsmm\ of two times the SM value would be obtained.

 It is also shown that at large values of both, \tb\ and the trilinear coupling,  negative interferences can lead to \Bbsmm\ values a factor three below the SM value, so even if values below the SM are found experimentally, this does not exclude supersymmetry, but constrains the parameter space.

\section{Acknowledgements} Support from the Deutsche Forschungsgemeinschaft (DFG) via a Mercator Professorship
(Prof. Kazakov) and the Graduiertenkolleg  "Hochenergiephysik und Teilchenastrophysik" in Karlsruhe  is
greatly appreciated. Furthermore, support from the Deutsche Luft und Raumfahrt (DLR) and the Bundesministerium
for Bildung und Forschung (BMBF) is acknowledged.

%\bibliographystyle{lucas_unsrt}
%\bibliography{mybib}
%\end{document}
\bigskip
\providecommand{\href}[2]{#2}\begingroup\raggedright\endgroup

\end{document}